# Electrically tunable circular photocurrent *via* local-field induced symmetry breaking at a metal-MoTe$_2$ interface

Butian Zhang[1,2,†] , Kexin Wang[3,†], Jun-Tao Ma[2], Yiya Guo[1,2], Chengyu Yan[1,2], Xin Yi[3],

Luojun Du[4,5], Youwei Zhang[1,2], Hua-Hua Fu[2,*], Shun Wang[1,2,*]

[1] National Gravitation Laboratory, MOE Key Laboratory of Fundamental Physical

Quantities Measurement, Huazhong University of Science and Technology, Wuhan

430074, People's Republic of China

[2] School of Physics, Huazhong University of Science and Technology, Wuhan

430074, People's Republic of China

[3] Kunming Institute of Physics, Kunming 650223, Yunnan Province, China

[4] Beijing National Laboratory for Condensed Matter Physics; Key Laboratory for

Nanoscale Physics and Devices, Institute of Physics, Chinese Academy of Sciences,

Beijing, 100190, China

[5] School of Physical Sciences, University of Chinese Academy of Sciences, Beijing

100190, China

ABSTRACT. Transition metal dichalcogenides (TMDCs) constitute a promising platform for symmetry-engineered responses to circularly polarized light. The high crystal symmetry of centrosymmetric 2H-phase TMDCs inherently forbids the circular photogalvanic effect, thereby necessitating external stimuli such as electric fields or strain to lower the symmetry for its activation. While Schottky junctions provide a ubiquitous built-in field for potentially inducing circular photocurrents, the mechanism for the generation and control of circular photocurrents in TMDCs is not understood. In this study, we fabricated a localized gold–$MoTe_2$ heterostructure and demonstrate a pronounced circular photocurrent at the interface under normal incidence. The photocurrent is attributed to circular photogalvanic effect governed by the strength and direction of the built-in electric field, enabling continuous modulation *via* an external bias. First-principles calculations show that the gold interface induces a spin splitting in the valence bands of $MoTe_2$, establishing a valley-dependent spin ordering. The observed circular photocurrent from multilayer 2H-$MoTe_2$ under normal incidence indicates the breaking of $C_3$ rotational symmetry by the local in-plane field. These results establish an effective strategy for developing voltage-tunable circularly polarized photodetectors and valleytronic devices.

**INTRODUCTION**

As fundamental electronic degrees of freedom beyond charge, spin and valley have emerged as promising platforms for next generation low-power electronics and quantum computing. [1-3] Efficient injection, manipulation, and detection of spin- and valley-polarized carriers in devices are essential for realizing their applications. In this context, the circularly polarized photocurrent (CPC), a light helicity-dependent photocurrent, serves as a powerful tool by enabling both the direct optical injection of spin/valley polarization and the probing of the underlying mechanisms. [4] In transition metal dichalcogenides (TMDCs), the inequivalent ±K valleys in the hexagonal Brillouin zone are well separated in momentum space, which suppresses intervalley scattering and provides robust information retention.[5] This separation is further exploited by circularly polarized light, which selectively excites individual K or -K valleys by angular momentum conservation. [6] When combined with the spin-orbit coupling in these materials, this selection rule further enables the generation of spin-polarized carriers at resonant wavelengths.

Crucially, converting such spin- and valley-polarized excitations into a measurable, directional CPC requires a net symmetry breaking that defines a preferred direction for carrier flow. This is commonly achieved by breaking the inversion symmetry, either through the intrinsic non-centrosymmetric crystal structure[7] or *via* engineered asymmetry in the device architecture. For centrosymmetric even-layer or bulk TMDCs, CPC has been achieved by lowering the material's symmetry through heterostructure engineering, [8] or by breaking the device symmetry *via* external fields such as non-uniform illumination, [9] mechanical strain, [10-12] or applied electric fields such as ionic gating. [13] Electric-field control has attracted wide attention due to its real-time and reversible tunability, enabling electrically programmable functionalities in devices such as reconfigurable polarized photodetectors and quantum information platforms.

The built-in electric field of a Schottky junction, naturally present at the metal contacts of many devices, has been identified as a possible origin of CPC. Initially, Dhara *et al.*[14] found CPC at the contact between metal and silicon nanowires arising from circular photogalvanic effect (CPGE), which was attributed to chirality generation by the local electric field—a process involving only orbital degrees of freedom without breaking spin degeneracy. The role of Schottky barrier in generating CPC has also been observed in two-dimensional material systems. In type-II Weyl semimetal TaIrTe$_4$[15] without inversion symmetry, the Schottky barrier was suggested to tilt the Fermi level near the Weyl nodes, leading to momentum-space-imbalanced optical transitions and thereby giving rise to CPGE. Although CPGE is formally forbidden by the crystal symmetry of most TMDCs (e.g., 2H, 1T′ phases), [16, 17] the built-in electric field of a Schottky barrier is expected to lift this symmetry constraint. A comparative study of monolayer MoSe$_2$ devices[18] shows that CPC at normal incidence emerges with direct metal contact but vanishes with an hBN tunnel barrier, establishing the critical role of Schottky contact electric fields. In multilayer 1T'-MoTe$_2$,[19] CPC under 4.0 μm excitation is found concentrated near the electrodes and was attributed to a third-order rather than second-order CPGE.

While a handful of studies have revealed the significant role of Schottky barriers in CPC generation within TMDCs, the underlying physical mechanism remains unclear. Key aspects that are not yet well understood include how the Schottky contact modifies the electronic band structure of TMDCs, the potential involvement of spin, valley, or orbital degrees of freedom, and the principles governing the magnitude and polarity of CPC under electrical control. Furthermore, the understanding of Schottky-barrier influence on CPC in multilayer 2H-phase TMDCs is still lacking, given their greater environmental stability compared to other phases and higher carrier mobility relative to monolayer counterparts. Additionally, the previously reported CPC signals are detected near the electrodes, where performance could be noncompatible and nonreproducible due to variations in fabrication processes and different work functions of adhesion metals (e.g., Ti/Au, Pd/Au).[20, 21] It is therefore essential to develop tailored device architectures that can isolate the local Schottky junction, enabling a clearer understanding of its specific role in the generation and modulation of CPC.

To improve the comparability of the devices in terms of contact quality and symmetry control, this study proposes to construct a gold (Au)–$MoTe_2$ heterostructure within the channel rather than at the electrodes. Multilayer 2H-phase $MoTe_2$ with a centrosymmetric structure is used to exclude the CPC arising from the intrinsic symmetry breaking. In the Au-$MoTe_2$ device, pronounced CPC with opposite signs are observed in the device when the laser spot was positioned at the left and right edges of the Au film, which is attributed to second-order CPGE. Meanwhile, this structure enables effective bias voltage control of the magnitude and sign of CPC, indicating that the direction of the built-in electric field can influence the polarity of the CPC. Based on first-principles calculations, we reveal the key role of the Au film in causing the spin splitting in multilayer $MoTe_2$. The observation of the CPC indicates a breaking of the $C_3$ rotation symmetry at the Au-$MoTe_2$ interface, which is driven by the in-plane local electric field. This study provides a new strategy for the design of electrically controlled CPC in TMDCs by exploiting their spin and valley degrees of freedoms.

**RESULTS AND DISCUSSION**

Figure 1(a) illustrates the device structure of the Au–$MoTe_2$ device, where multilayer 2H-$MoTe_2$ is dry-transferred onto a pre-patterned Au film. The 2H-$MoTe_2$ flakes used in this study have thicknesses ranging from 20 to 60 nm. The Au film thickness is approximately 14 nm, as determined by atomic force microscopy (AFM) (Figure S1). Ti/Au contacts were deposited as source/drain electrodes to ensure low contact resistance. A 1100 nm laser was focused through a 50× objective, producing a Gaussian spot (FWHM ≈ 12 μm) for normal-incidence CPC measurements.

The normal-incidence CPC response was compared between the Au–$MoTe_2$ device (Figure 1(b)) and a conventional full-channel $MoTe_2$ device (Figure 1(c)) by recording the photocurrent while rotating a quarter-wave plate. The angle-dependent total photocurrent, $I_{PC}$, was fitted using the expression:

$$I_{PC} = C\sin(2\theta) + L\sin(4\theta + \varphi) + I_0$$

where $C$ and $L$ represent the circularly and linearly polarized components of the photocurrent, respectively, $\varphi$ is a phase offset that depends on the angle between the crystal lattice orientation and the initial polarization axis set by the linear polarizer, and $I_0$ is the polarization-independent component.

The fitting results of $C$ at different drain-source voltages $V_\text{d}$ are summarized in Figure 1(d). The CPC from conventional MoTe₂ device remains nearly zero and shows weak dependence on $V_\text{d}$ at low voltages, as shown in the inset. In contrast, the CPC in the Au–MoTe₂ device exhibits a near-linear and monotonic increase in amplitude with $V_\text{d}$, reversing its sign at approximately –3 mV. This demonstrates effective electrical control over both the magnitude and polarity of the CPC. This tunability is directly evidenced in the transfer characteristics (Figure S2), which show that the photocurrent difference between left- and right-handed illumination depends on both the $V_\text{d}$ and gate ($V_\text{g}$) voltages. The observed asymmetry in the CPC–$V_\text{d}$ curve—the distinct slopes in the positive and negative bias regions—could stem from the asymmetric device geometry, primarily the off-center position of the Au film and the inherent differences in the local Schottky barriers at the two junction edges. A direct comparison of the photocurrent composition at zero bias ($V_\text{d}$= 0 V) further highlights the significant role of the Au film. When the spot is positioned on the Au–MoTe₂ heterostructure, the photocurrent shows a significant π-periodic modulation, signifying a strong helicity-dependent response (Figure 1(e)). Conversely, when the laser spot illuminates the conventional device, the photocurrent $I_{PC}$ exhibits a dominant π/2-period variation with $\theta$, characteristic of a linear-polarization related effect (Figure 1(f)), indicating a negligible CPC, which is consistent with previous observations in centrosymmetric TMDCs.[9, 22] In the Au–MoTe2 structure, the emergence of a substantial CPC even at zero bias highlights the intrinsic ability of the metal–semiconductor interface to generate CPC through its built-in symmetry breaking.

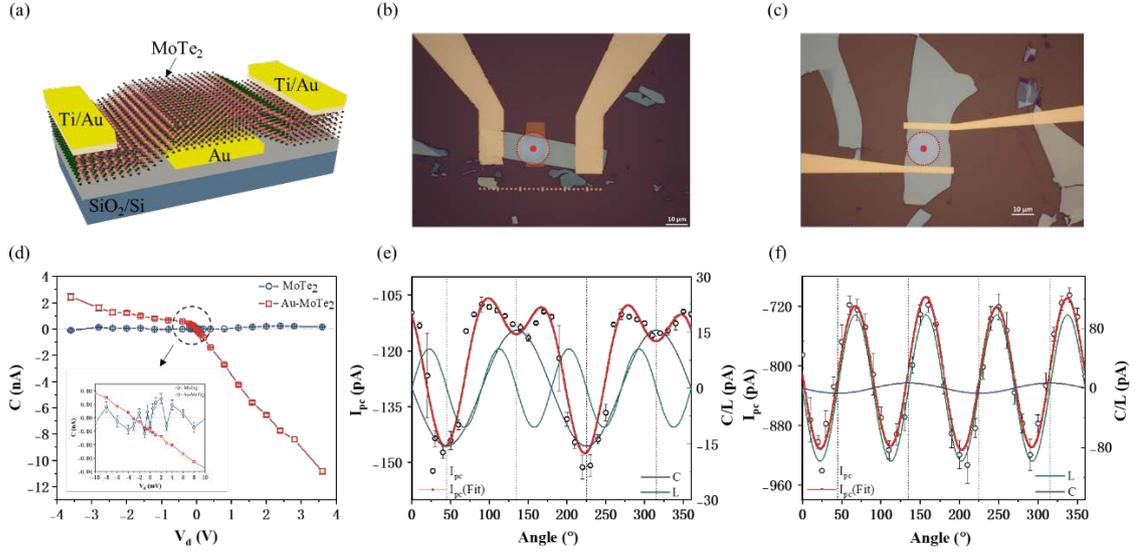

**Figure 1.** (a) Schematic diagram of the Au–MoTe₂ device structure. Optical image of the Au-MoTe₂ device (b) and the conventional MoTe₂ device (c). The red dot and circle mark the center and full width at half maximum of the laser spot, respectively. (d) CPC as a function of drain voltages ($V_\mathrm{d}$) for both devices. The error bars represent the standard error of the fit. (e, f) Total photocurrent $I_\mathrm{pc}$ and the fitted circular ($C$) and linear ($L$) polarization-dependent photocurrents versus the quarter-waveplate angle for the (e) Au-MoTe₂ and (f) MoTe₂ at $V_\mathrm{d} = 0$ V. All data were acquired at a gate voltage ($V_\mathrm{g}$) of 0 V. Error bars in (d) denote the standard error of the fit, while those in (e, f) indicate the standard deviation from repeated measurements.

To spatially resolve the origin of the circularly polarized photocurrent (CPC) at the Au–MoTe2 interface, we characterized a device based on a larger MoTe2 flake underlaid by a ~30 µm-wide Au film. As shown in Figure 2(a), CPC signals were separately measured at the left and right edges of the interface region, with non-overlapping laser spots. Figure 2(b) plots the bias dependence of the CPC at these two locations. Under zero external bias, the CPC values at the two edges exhibit opposite signs, which is further corroborated by the distinct and reversed circular polarization dependences of Figure 2 (c, d). For both edges, the CPC depends nearly linearly on $V_{d}$ and vanishes at specific bias voltages. Notably, the values of $V_{d}$ required to quench the CPC are of opposite polarity at the left and right edges. These observations confirm that the magnitude and sign of the CPC are governed by the combined action of the built-in and externally applied electric fields. The applied $V_{d}$ not only modulates the local built-in fields but also enhances the dissociation of photogenerated excitons and accelerates carrier drift, thereby amplifying the CPC signal. It simultaneously influences the magnitude and direction of the polarization-independent photocurrent $I_{0}$, primarily through the photovoltaic effect. Notably, the CPC and $I_{0}$ reverse their signs at different $V_{d}$ in Figure 2(b), indicating distinct physical origins. This clear separation in their electrical tuning behaviors provides strong evidence that the observed CPC originates from the intrinsic response rather than from any artifacts in the background photocurrent.

Figure 2(e) illustrates the schematic band diagrams of the device under different bias voltages, providing a qualitative explanation of the mechanism by which $V_\text{d}$ modulates the CPC. Two junction regions (J1 and J2) form at the opposite edges of the Au–MoTe$_2$ interface. At $V_\text{d} = 0$ V, their built-in electric fields point in opposite directions, accounting for the observed CPC with opposite signs. Under a positive $V_\text{d}$, J1 becomes reverse-biased, leading to increased band bending and an extended built-in electric field. Meanwhile, the forward bias at J2 leads to a flattening of the band profile and a reduction in the built-in electric field. The resulting contrast in local fields explains the opposite trends in CPC magnitude at the two edges: it increases at J1 but decreases at J2 with rising $V_\text{d}$. As $V_\text{d}$ increases further, the band bending at J2 eventually reverses, leading to a corresponding reversal in CPC polarity. A symmetric behavior occurs under negative $V_\text{d}$, with the roles of J1 and J2 interchanged. These results demonstrate that $V_\text{d}$ controls the CPC by directly modulating the strength and direction of the local built-in fields.

When the electric fields at J1 and J2 become comparable, the overall band profile becomes symmetric and the total CPC by summarizing the value at J1 and J2 should vanish. Referring back to the experimental results in Figure 1(d) and (e), we note that a weak CPC can still be observed at the Au–MoTe$_2$ contact region even at $V_\text{d} = 0$ V.

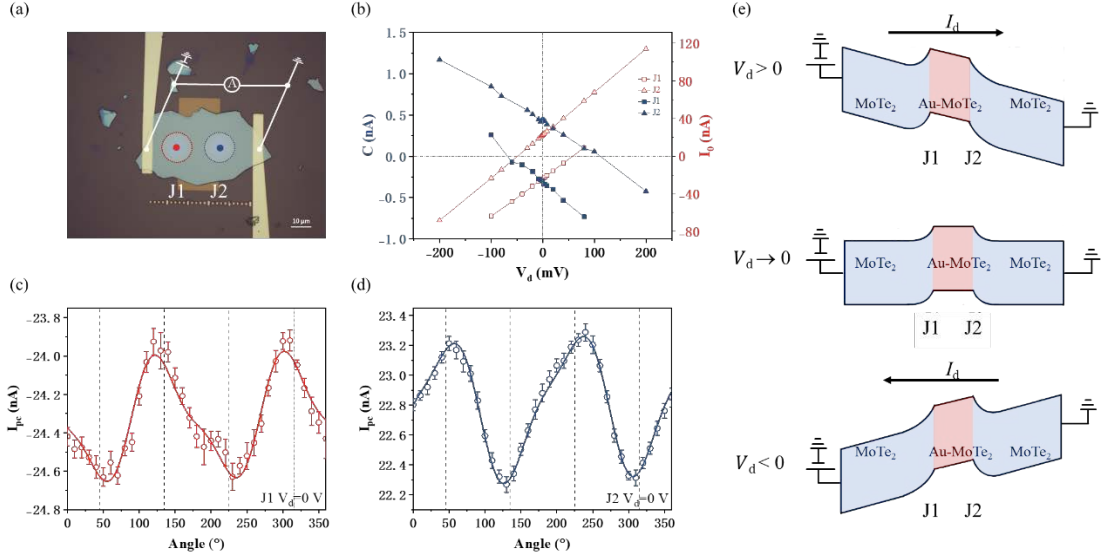

Figure **2**. (a) Optical image of the Au–MoTe₂ device, showing the laser spot positions at the left (J1) and right (J2) edges of the heterointerface. (b) CPC (navy solid symbols) and $I_0$ (red hollow symbols) obtained from two edges as a function of $V_d$ at $V_g = 0\ V$. (c, d) Total photocurrent $I_{PC}$ as a function of the quarter-wave plate angle with the laser positioned at the (c) left and (d) right edges, respectively, measured at $V_d = V_g = 0$ V. Error bars denote the standard deviation from repeated measurements. (e) Schematic band diagrams illustrating the proposed electronic structures at the two junctions under different $V_d$.

This indicates that the CPC at the two junctions is not fully balanced, which can originate from intrinsic differences in the local Schottky barriers and/or a slight off-center laser spot position that unequally excites the two junctions. The CPC vanishes completely at $V_d = -3$ mV, a condition which achieves the compensation of the net CPC. This occurs when the applied bias tunes the band bending and barrier heights to a state where the contributions from the two junctions cancel each other out, despite any inherent asymmetry or uneven optical excitation. While $V_d$ directly tunes the interfacial band bending, in contrast, the gate voltage $V_g$ modulates the CPC through a different pathway (Figure S3). Because the MoTe₂ channel is partially in contact with the underlying Au film, the metallic screening effect confines the influence of $V_g$ primarily to the region of MoTe₂ in contact with the SiO₂ substrate, where it modulates the carrier concentration. The resulting carrier concentration gradient drives a diffusion current, which effectively mimics an in-plane electric field. This diffusion current can either counteract or reinforce the externally applied $V_d$, thereby producing the local extrema in CPC observed at certain gate voltages.[23]

Our previous study has shown that introducing a spatially nonuniform light spot in specially designed electrode structures can break the system symmetry, leading to the observation of a circulating CPC, which can be explained as the inverse spin Hall effect (ISHE).[9] However, this mechanism necessitates a highly asymmetric geometry between the light spot and the electrode, which is absent here. Furthermore, the observation of a clear CPC signal even under illumination with a large laser spot (4× objective, spot size much larger than the device channel) as shown in Figure S4, confirms that the effect is independent of spatial nonuniformity of the light spot, which is a key requirement for the ISHE mechanism. More decisively, in the electrode structure used in this study, no CPC was observed when a Gaussian light spot illuminated MoTe₂ devices without an Au substrate, indicating that ISHE is not the primary contributor. This also demonstrates that the pristine, unperturbed material itself lacks a sufficient intrinsic Berry curvature to generate an observable CPC signal. More importantly, the conventional ISHE or valley Hall effects (VHE) would generate a transverse charge current, perpendicular to the direction of the driving electric field. [24-26] It is worth noting that, in principle, the response tensors for the ISHE in low-symmetry systems may exhibit longitudinal components. However, such a response has not yet been observed in TMDCs, which is likely due to their low spin Hall conductivity. [27] In our device geometry, the observed CPC is collected longitudinally along the channel, parallel to the applied drain-source or built-in field, which rules out the conventional VHE or ISHE. Furthermore, the normal-incidence condition precludes contributions from the circular photon drag effect (CPDE). Among possible CPC mechanisms, the observed phenomena are consistent with CPGE.

To elucidate the energy band origin of the CPC, we investigated its wavelength dependence. As shown in Figure 3 (a, b), the CPC and $I_0$ peak at incident wavelengths of 1100 nm and 1200 nm, respectively, indicating that the CPC primarily originates from the A excitons in the K and K′ valleys of MoTe$_2$. Shorter wavelength to 900 nm will lead to a further reduction in CPC due to the overlap and partial cancellation of contributions from A exciton and B exciton with opposite spin splitting. This interpretation aligns with the observation of CPC near the metal electrode of MoSe$_2$ devices, [9] reinforcing that the Schottky barrier-induced CPC in TMDCs stems predominantly from the K point. Moreover, the previously reported third-order CPGE in 1T'-MoTe₂[19] is independent of the excitation photon energy, whereas in our experiments, the A exciton in 2H–MoTe₂ at the Au–MoTe₂ interface was identified as the main contributor to the CPC, indicating that the third-order CPGE mechanism does not apply to this study. The second-order CPGE could origin from K and K' valleys, which is usually associated with the non-vanishing Berry curvatures and should be zero for intrinsic 1T' and bulk 2H-phase TMDCs. The peak wavelength of $I_0$ at 1200 nm generally corresponds to the wavelength of maximum light absorption. The difference in peak wavelengths between CPC and $I_0$ therefore suggests that the spin and valley polarization ratio is higher at 1100 nm. This interpretation is plausible within the framework of the spin-splitting bands presented in Figure 4.

For our sample, both CPC and $I_0$ rise as the incident light intensity increases (Figure S5). This correlation suggests that the CPC is closely linked to the efficiency of carrier separation and collection. In TMDCs, the background photocurrent mainly arises from the photovoltaic effect (PVE) and the photothermoelectric effect (PTE). As shown in Figure 3(c), the dominant mechanism can be identified by the power-law relation $I \propto P^\alpha$ where $\alpha \approx 1$ corresponds to the PVE and $\alpha \approx 2/3$ to the PTE. The measured $\alpha \approx 1.02$ from the MoTe₂ region indicates PVE dominance, while $\alpha \approx 0.74$ from the Au–MoTe₂ region suggests a mixed contribution from both PVE and PTE. The PTE arises from a laser-induced temperature gradient at an interface between materials with different Seebeck coefficients, which generates a thermoelectric voltage and drives a corresponding current. In the Au–MoTe₂ heterostructure, laser absorption by the partially overlaid metallic Au film creates a pronounced non-uniform temperature profile across the MoTe$_2$ channel, leading to the emergence of PTE.

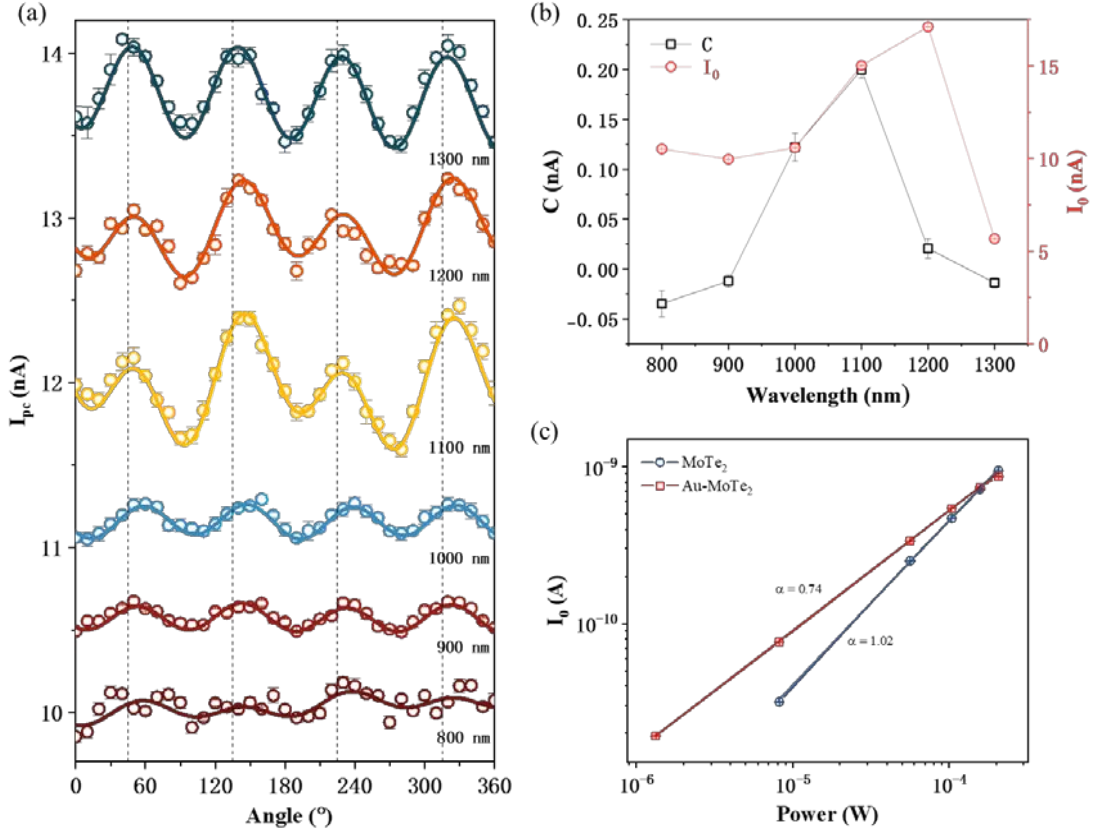

**Figure 3.** (a) Wavelength-dependent total photocurrent $I_{pc}$ of the Au–MoTe$_2$ device as a function of the quarter-waveplate angle, measured at $V_d = V_g = 0$ V. Error bars represent the standard deviation across multiple measurements. The curves have been vertically offset for clarity. (b) C component and $I_0$ component as a function of incident wavelength. Error bars indicate the standard error from the curve fitting. (c) Log–log plot of $I_0$ versus incident laser power for both the Au–MoTe$_2$ and the pristine MoTe$_2$ region.

Currently, the exact mechanism by which Schottky contacts induce CPC in centrosymmetric 2H-TMDCs remains unclear. To elucidate the microscopic mechanism, we performed first-principles calculations on a bilayer MoTe$_2$ matched with an Au (111) surface, as shown in Figure 4a. Figure 4b and 4c present the band structures of pristine bilayer MoTe$_2$ and the Au–MoTe$_2$ heterostructure, respectively. The Au interface induces an additional energy band that spans across the MoTe$_2$ band gap, connecting its valence and conduction bands. More crucially, it induces an additional spin splitting in the MoTe$_2$ valence bands (for both VB1 and VB2) at the K and K' point, beyond the intrinsic Ising spin–orbit coupling. Notably, a similar splitting is reproduced by applying a strong out-of-plane electric field to pristine MoTe$_2$ (Figure 4d). This indicates that the interfacial effect of Au is electronically equivalent to applying such a field, which breaks the spatial inversion symmetry and creates a potential difference between the two MoTe$_2$ layers. This calculated symmetry-breaking scenario provides a microscopic basis for our proposed model of CPC generation. In centrosymmetric bulk or multilayer MoTe$_2$, the symmetry-breaking field predominantly affects the bilayer or few-layer closest to the metal, which corresponds to the region where the Schottky built-in field is strongest. It is proposed to selectively lift this cancellation by inducing opposite spin splitting in the K and K' valleys—e.g., raising spin-up in one valley while raising spin-down in the other—thus creating a valley-dependent spin ordering. In intrinsic centrosymmetric bilayer and bulk 2H-phase TMDCs, the local inversion symmetry breaking gives rise to "hidden" spin polarization[28] and allows the coupled layer–valley degrees of freedom to be optically accessed.[29] Similarly, within this interfacial bilayer and few-layer, the layer-localized valleys largely retain their optical selection rules due to incomplete interlayer orbital hybridization. Therefore, at a given central wavelength, circularly polarized light preferentially excites carriers from one valley over the other, as the resonance condition dominates the absorption. This leads to a population where a majority of carriers exhibit a specific valley and spin polarization.

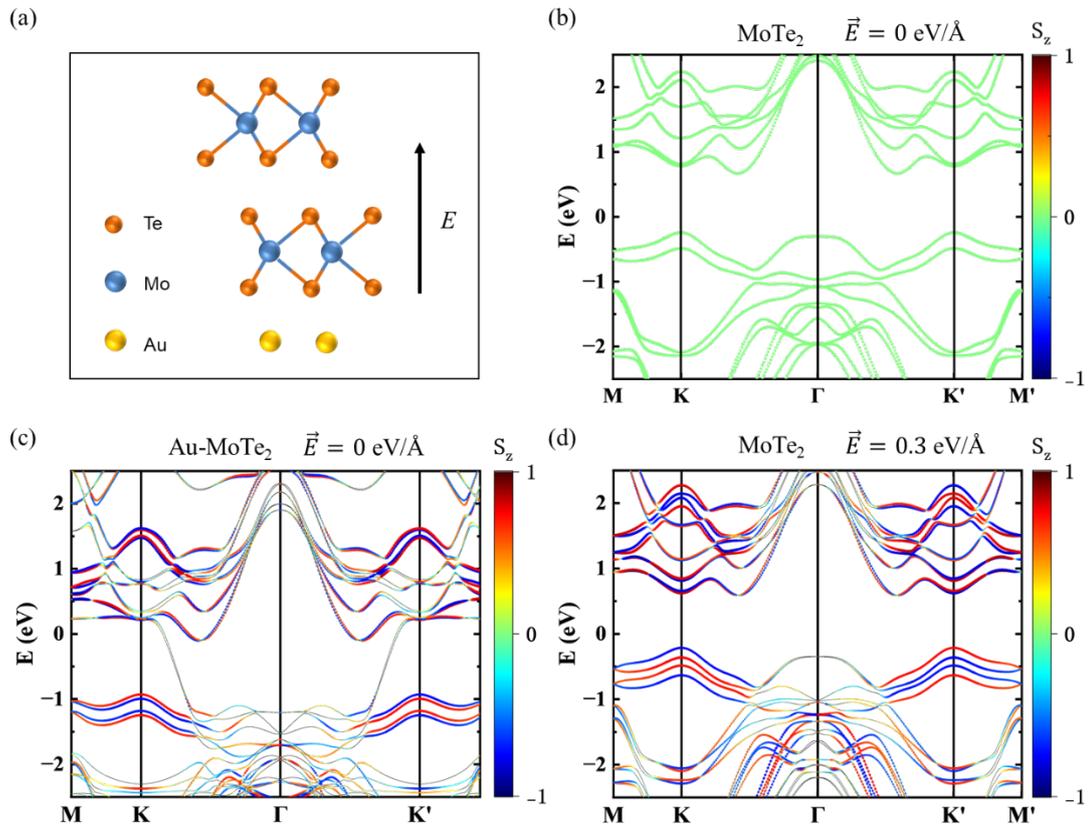

**Figure 4.** (a) Atomic structure of the Au–MoTe₂ heterojunction used in first-principles calculations. (b–d) Corresponding electronic band structures: (b) pristine MoTe₂, (c) Au-MoTe₂ heterojunction without an external field, and (d) pristine MoTe₂ under an out-of-plane electric field of 0.3 V/Å.

Such a photoinduced valley and spin polarization can also occur in monolayer materials (with intrinsic inversion symmetry breaking) or in ionic-gated bilayer systems. However, in monolayer materials, the presence of a horizontal mirror plane ($\sigma_h$) within the $D_{3h}$ point group strictly forbids any non-zero CPGE tensor elements from a group-theoretical perspective, preventing the generation of a macroscopic CPGE current under normal incidence. In ionic-gated bilayer systems such as $WSe_2$, although a vertical electric field can break the inversion symmetry—reducing the system's symmetry to $C_{3v}$ and allowing non-zero CPGE tensor elements—the retained in-plane rotational symmetry results in a negligibly small photocurrent.[13] Such systems typically rely on oblique incidence (introducing a vertical electric field component of the light) or an in-plane bias to introduce the necessary directional symmetry breaking. In contrast, the Au–$MoTe_2$ interface in this work operates differently. With the inversion symmetry broken by the Au film, the intrinsic asymmetry of the local built-in electric field further breaks the $C_3$ symmetry—which meets the requirement to induce CPGE.[30] Consequently, under the normal incidence, the system can directly generate and output a strong macroscopic CPGE current without relying on oblique illumination or any externally applied in-plane bias. To exclude the possibility that the observed CPGE originates from substrate-induced strain, we replaced the Au substrate with $Al_2O_3$ and measured the photocurrent at both edges of the heterojunction (Fig. S6). No significant circular photocurrent was observed at either edge, confirming that strain alone is insufficient to activate CPGE in this system.

**CONCLUSIONS**

In this work, we demonstrate that the Schottky barrier at an Au–MoTe$_2$ interface can generate and electrically control the CPC in centrosymmetric multilayer MoTe$_2$ under normal incidence. Spatially resolved photocurrent measurements confirm that CPC is exclusively activated at the interface, where the strength and direction of the local built-in electric field respectively govern the amplitude and sign of the CPC. This control is electrically tunable *via* an external bias, which modulates the Schottky barrier to enhance carrier separation and reverse the CPC polarity. The emergence of CPC is attributed to second-order CPGE by combining the current direction, the normal incidence condition, together with the wavelength dependence. First-principles calculations attribute this to the Au-induced spin splitting in the MoTe$_2$ valence bands at the K and K′ valleys, which lifts the spin degeneracy and creates a valley-dependent spin ordering under the interfacial field. This establishes the microscopic origin of the CPGE, wherein circularly polarized excitation results in an asymmetric momentum distribution of spin-polarized carriers. Different from previous observations on monolayer and ionic gated bilayer TMDCs, the C$_3$ symmetry of the multilayer MoTe$_2$ requires breaking by the local in-plane electric field at the Schottky interfaces. Our findings highlight the pivotal role of engineered Schottky interfaces in activating CPC in centrosymmetric 2D semiconductors and offer a viable device strategy for achieving voltage-controlled CPGE responses. Future work on CPC and other circular-polarization-dependent responses should carefully consider the potential contributions from Schottky barriers at the relevant metal–semiconductor interfaces.

ASSOCIATED CONTENT

**Supporting Information Available:** The following files are available free of charge.

Methods of device fabrication and measurements; Atomic Force Microscopy (AFM) characterization of the Au–MoTe$_2$ device; electrical characteristics of Au-MoTe$_2$ device under circularly polarized light illumination; Gate voltage dependence of the circular photocurrent; Angular dependence of photocurrent measured under 4× objective lens illumination (PDF)

**AUTHOR INFORMATION**

**\*Corresponding Authors**

Hua-Hua Fu − School of Physics, Huazhong University of Science and Technology, Wuhan 430074, People's Republic of China; orcid.org/0000-0003-3920-6324; Email: hhfu@hust.edu.cn

Shun Wang − National Gravitation Laboratory, MOE Key Laboratory of Fundamental Physical Quantities Measurement, and School of Physics, Huazhong University of Science and Technology, Wuhan 430074, People's Republic of China; orcid.org/0000-0001-6890-8371; Email: shun@hust.edu.cn

**Author Contributions**

Shun Wang and Butian Zhang conceived and designed the research. The experiments were designed and performed by Kexin Wang and Yiya Guo. Juntao Ma and Huahua Fu performed the DFT calculations. Youwei Zhang and Xin Yi assisted in the experimental improvements. Butian Zhang and Kexin Wang analyzed the data and wrote the manuscript with the assistance of Chengyu Yan, Luojun Du, and Shun Wang. Shun Wang and Hua-Hua Fu supervised the research and contributed to manuscript review & editing. All authors discussed the results and participated in revising the manuscript. †These authors contributed equally to this work.

**ACKNOWLEDGMENT**

S.W. acknowledges the support from the National Key R&D Program of China (No. 2021YFC2202300). B.Z. acknowledges the support from the Experimental Technology Research Project (No. 2025a-1.1-5), Lab and Equipment Administration Department, Huazhong University of Science and Technology.